\documentclass[conference]{IEEEtran}
\IEEEoverridecommandlockouts

\usepackage{comment}
\usepackage{cite}
\usepackage{physics}
\usepackage{caption} 
\usepackage{amsmath,amssymb,amsfonts}
\usepackage{mathtools}
\usepackage{algorithmic}
\usepackage{graphicx}
\usepackage{textcomp}
\usepackage{tikz}
\usetikzlibrary{quantikz}
\usepackage{algorithm}
\usepackage{algorithmic}
\usepackage{float}
\usepackage{subcaption}
\usepackage[T1]{fontenc}

\usepackage{hyperref}

\def\BibTeX{{\rm B\kern-.05em{\sc i\kern-.025em b}\kern-.08em
    T\kern-.1667em\lower.7ex\hbox{E}\kern-.125emX}}

\begin{document}

\def\myvdots{\ \vdots\ }
\title{Quantum Internet: from Medium Access Control to Entanglement Access Control}

\makeatletter
\newcommand{\linebreakand}{%
  \end{@IEEEauthorhalign}
  \hfill\mbox{}\par
  \mbox{}\hfill\begin{@IEEEauthorhalign}
}
\makeatother

\author{
    \IEEEauthorblockN{Jessica Illiano\IEEEauthorrefmark{1}, Michele Viscardi\IEEEauthorrefmark{1}, Seid Koudia\IEEEauthorrefmark{1}, Marcello Caleffi \IEEEauthorrefmark{1}\IEEEauthorrefmark{2} Angela Sara Cacciapuoti\IEEEauthorrefmark{1}\IEEEauthorrefmark{2}} \IEEEauthorblockA{\IEEEauthorrefmark{1}
        \href{www.quantuminternet.it}{www.QuantumInternet.it} research group, Department of Electrical Engineering and Information Technologies\\
        University of Naples \textit{Federico II}, Naples, 80125, Italy\\
        \href{mailto:jessica.illiano@unina.it}{jessica.illiano@unina.it},
        \href{mailto:mi.viscardi@studenti.unina.it}{mi.viscardi@studenti.unina.it}\\\href{mailto:seid.koudia@unina.it}{seid.koudia@unina.it}, 
        \href{mailto:marcello.caleffi@unina.it}{marcello.caleffi@unina.it},
        \href{mailto:angelasara.cacciapuoti@unina.it}{angelasara.cacciapuoti@unina.it}\\
    }
    \IEEEauthorblockA{\IEEEauthorrefmark{2}National Inter-University Consortium for Telecommunications (CNIT), Naples, 80126, Italy}
}


\maketitle

\begin{abstract}
Multipartite entanglement plays a crucial role for the design of the Quantum Internet, due to its potentiality of significantly increasing the network performance. In this paper, we design an \textit{entanglement access control} protocol for multipartite state, which exhibits several attractive features. Specifically, the designed protocol is able to \textit{jointly} extract in a \textit{distributed} way an EPR pair from the original multipartite entangled state shared by the set of network nodes, and to univocally determines the identities of the transmitter node and the receiver node in charge of using the extracted EPR pair. Furthermore, the protocol avoids to delegate the signaling arising with entanglement access control to the classical network, with the exception of the unavoidable classical communications needed for EPR extraction and qubit teleportation. Finally, the protocol supports the \textit{anonymity} of the entanglement accessing nodes.
    
\end{abstract}


\section{Introduction}
\label{sec:1}
 
\begin{figure*}[t!]
\centering
    \begin{subfigure}[c]{0.48\textwidth}
        \centering
        \includegraphics[width=\textwidth]{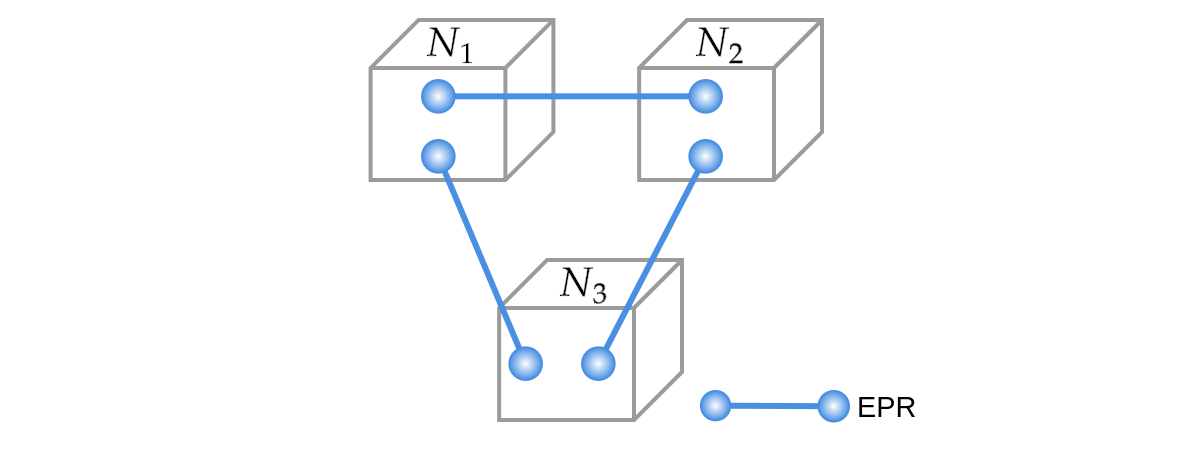}
        \caption{EPR-based connectivity. For directly interconnecting three nodes, a minimum of three EPR pairs must be generated and distributed through the network. Once distributed, each EPR establishes a virtual quantum link between a fixed pair of nodes, namely, the nodes storing a member of the EPR pair.}
        \label{fig:01.a}            
    \end{subfigure}
    \hfill
    \begin{subfigure}[c]{0.48\textwidth}
        \centering
    	\includegraphics[width=\textwidth]{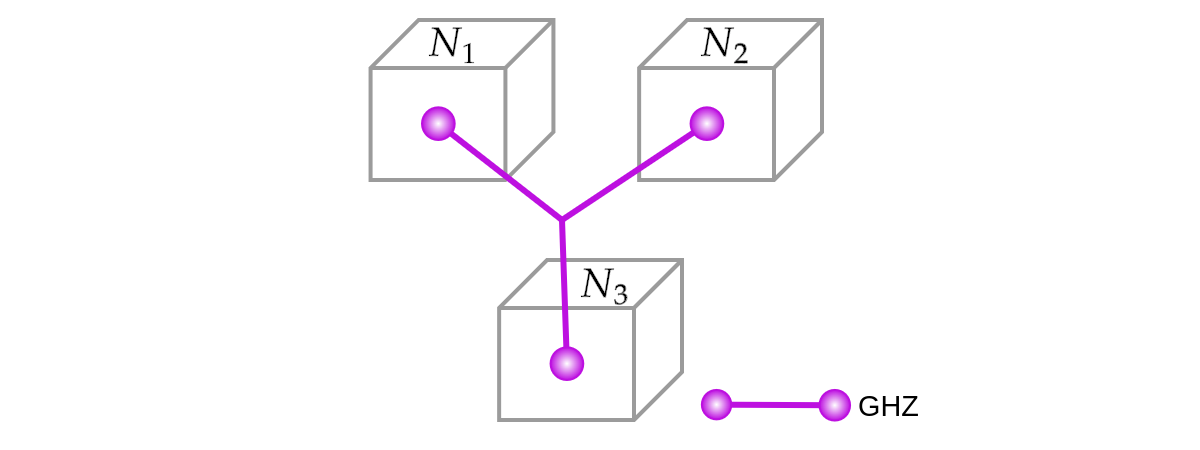}
    	\caption{Multipartite-based connectivity. A direct connection between any pair of nodes can be obtained by distributing a multipartite state -- i.e., a GHZ state -- among the network nodes. In fact, an EPR between any pair of nodes can be extracted from a multipartite state on-demand.}
    	\label{fig:01.b}                   
    \end{subfigure}
    \caption{\textit{A-priori} vs. \textit{on-demand} connectivity.}
    \label{fig:01}
    \hrulefill
\end{figure*}
 
Recently, the research community is devoting increasing attention to the role of multipartite entanglement into the design of the Quantum Internet \cite{CalCacBia-18,CacCalTaf-20,PirBra-16,KozWehVan-22}, due to its potentiality to significantly increase the network performance \cite{RamPirDur-21}.

Indeed, multipartite entanglement ensures not only privacy and anonymity \cite{KhaKhaReh-22}, but it also enables the so-called \textit{on-demand} connectivity \cite{IllCalMan-22}. Specifically, for the sake of exemplification, let us consider three nodes, say nodes $N_1$, $N_2$ and $N_3$ as in Figure~\ref{fig:01.a}. In EPR-based networks \cite{VanTou-13,Weh-19,PomDonWeh-21}, to establish a direct\footnote{The term ``direct connectivity'' refers to the availability of an EPR pair shared between two nodes (regardless of how this EPR has been distributed to the nodes, i.e., through direct quantum link or multi-hop quantum path), without the need of any additional \textit{helper} such as an intermediate third node implementing entanglement swapping. Accordingly, the EPR pair can be straightly exploited -- as instance, for quantum teleportation -- by the two nodes without involving any additional node.} connectivity among any pair of the set, three EPR pairs must be properly shared by the nodes. Specifically, the EPRs must be distributed so that each node -- say $N_1$ -- stores two qubits from two different EPR pairs, with each EPR pair shared with a different node -- i.e, $N_2$ and $N_3$, respectively. Accordingly, the identity of the nodes that can exploit entanglement as a \textit{communication resource} is fixed a-priori, with no possibility of adapting to time-varying communication needs.
 
Differently, by considering multipartite-based networks \cite{RamPirDur-21}, multipartite entangled states --  such as GHZ states \cite{GreHorZei-89,GreHorZei-90,IllCalMan-22} -- enable the extraction of an EPR pair between any pair of nodes \textit{at run-time}, depending on the instantaneous communication needs. This key feature enables full connectivity among $n$ nodes, without (unreasonably) requiring a number of communication qubits\footnote{As a matter of fact, a fair comparison between the two strategies requires an identical amount of resources -- i.e., communication qubits -- at each node.} at each node scaling with $\mathcal{O}(n)$. With reference to the example of Figure~\ref{fig:01.a}, by distributing a 3-qubit GHZ state through the network so that each node stores one qubit of the state as in Figure~\ref{fig:01.b}, an EPR pair can be extracted and shared by any pair of nodes, with the identities of the entangled nodes chosen at run-time.

Clearly, this key feature of on-demand connectivity requires proper management and coordination among the entangled nodes, since they all \textit{share} the same multipartite state and yet only one EPR pair can be extracted.

Remarkably, the \textit{entanglement access control} functionality is not required only in multipartite-based networks but it must be available also in EPR-based networks\footnote{Hence, the design of an entanglement access control protocol is mandatory regardless of the bipartite or multipartite nature of the entanglement.} \cite{IllCalMan-22}. In fact, entanglement as a communication resource is \textit{blind} with respect to the role subsequently assumed by the node in the exploitation of the resource, i.e., ``transmitter'' or ``receiver''. Indeed, without an entanglement access control, two nodes sharing an EPR could both try to use the resource as transmitter. As instance, the same resource -- e.g., the EPR pair in Figure~\ref{fig:01.a} between nodes $N_1$ and $N_2$ -- can be equivalently exploited for quantum communications from $N_1$ to $N_2$ or vice-versa. And, as a consequence, an entanglement contention problem arises, requiring a proper cooperation between $N_1$ and $N_2$. Additionally, when it comes to multipartite entangled states, \textit{entanglement access control} assumes an even more crucial role, due to the increased number of nodes that must coordinate for the EPR extraction operation.

Despite this, entanglement access control has been poorly investigated so far, by implicitly delegating it to some form of classical signaling through classical networks. This in turn would require an effective classical-quantum interface between the classical Internet and the Quantum Internet protocol stack, which is a still an open issue \cite{IllCalMan-22}. As a consequence, limiting classical signaling represents both a viable strategy accordingly to the current state-of-the-art as well as an interesting and yet-to-be solved research problem.

With this in mind, in this paper, we design an entanglement access control protocol with the objective of incorporating any signaling arising with entanglement access control into quantum communication resources, by exploiting the unconventional features of multipartite entanglement.

\section{Preliminaries}
\label{sec:2}
In the following, we first describe the system model and then we introduce the problem statement for the design of the entanglement access control protocol.

\subsection{System Model}
\label{sec:2.1}

\begin{figure*}[t!]
    \centering
    \includegraphics[width=\linewidth]{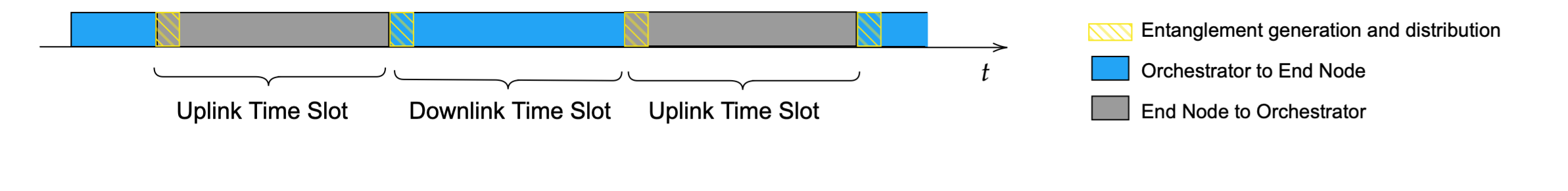}
    \caption{Graphical representation of the slotted time organization, with uplink -- namely, end-nodes to orchestrator -- transmissions interleaved with downlink -- namely, orchestrator to end-nodes -- ones.}
    \label{fig:02}
    \hrulefill
\end{figure*}     
    
For the sake of simplicity, we consider a star-network topology\footnote{This assumption is not restrictive since our protocol can be easily extended to peer-to-peer topologies. 
}, with a central node -- say $N_0$ -- referred to as the \textit{orchestrator} node and $n$ end-nodes -- $\{N_i\}_{i=1}^{n}$ -- interacting in a time-slotted fashion, according to a model reminiscent of IEEE 802.15.4. Specifically, the orchestrator plays the same role of a classical IEEE 802.11 access point, namely, it acts as some sort of gateway for communications outside the quantum network and it manages the communications of the nodes belonging to the network. The assumption of a super-node -- the orchestrator -- is quite common in the quantum literature \cite{AviRozWeh-22}, due to the complex mechanisms and the dedicated equipment underlying the entanglement generation and distribution process.

Orchestrator and end-nodes share a multipartite entangled state, and the overall goal of the entanglement access control protocol is to solve the entanglement contention problem among the nodes, with -- as final outcome -- the extraction of an EPR pair shared between a selected transmitter-receiver pair, by avoiding to delegate the access signaling to the classical network, as detailed in the following.

According to the time model represented in Figure~\ref{fig:02}, during the downlink, the orchestrator acts as transmitter and one of the end-nodes acts as receiver. The opposite holds during the uplink time-slot. It is worthwhile to note that, even if the adopted time-slotting automatically sets the transmitter node identity for downlink communications and the receiver node identity for the uplink ones, a contention of the entangled resource still occurs as discussed in the following.  

\subsection{Problem Statement}
\label{sec:2.2}
As previously mentioned, the design of the entanglement access control protocol aims at solving the entanglement contention  among the nodes, with the final result of sharing an EPR pair among the selected node pairs. More into detail, the design of the protocol is carried out by full-filling the following requirements: 
\begin{itemize}
    \item the protocol must be able to \textit{jointly} i) extract in a \textit{distributed} way an EPR pair from the original multipartite entangled state shared by the set of network nodes, and ii) univocally determine the identities of the node acting a transmitter and the one acting as receiver for the extracted EPR pair;
    \item the protocol abstains from delegating the signaling arising with entanglement access control to the classical network, with the exception of the unavoidable classical communications needed to i) extract an EPR pair from a multipartite entangled state, and ii) perform the teleportation protocol;
    \item the protocol supports the \textit{anonymity} of the selected nodes, i.e., transmitter and receiver identities should be disclosed only to each others, and they should be kept hidden to the remaining nodes.
\end{itemize}

From the above, it results that the protocol must be able not only to determine the identities of the two nodes eventually sharing the EPR pair, but also to associate a role -- transmitter or receiver -- to them. Mathematically, this can be modeled by introducing a proper functional $\chi(\cdot)$.

Specifically, given a set of nodes $\mathcal{N}=\{N_i\}^n_{i=0}$ and the set of roles $\mathcal{R}=\{R_t,\overline{R_t}\}$ -- with $R_t$ denoting the transmitter role and $\overline{R_t}$ the receiver role -- let us define the set $\mathcal{N_R}$ constituted by the ordered pairs of network nodes, where the first node of the pair acts as transmitter and the second one as receiver:
\begin{align}
    \label{eq:1}
    \mathcal{N_R}= \big\{ \, &  \{(N_i, N_j)\}^n_{i,j=0, i \neq j} : \nonumber \\
    & \quad N_i \text{ has role } R_t \, \wedge \, N_j \text{ has role }\overline{R}_t \big\}
\end{align}

Stemming from \eqref{eq:1}, the functional $\chi(\cdot)$ must be able to univocally and jointly determine the node pair sharing the EPR along with the roles associated to each element of the node pair, as follows:
\begin{equation}
    \label{eq:2}
    \chi : \, \mathcal{N_R} \rightarrow \{0,1\} \quad \textrm{ s.t. } \quad \sum_{i,j} \chi (N_i,N_j) = 1
\end{equation}

It should be noted that the joint requirements of distributed strategy and avoiding to delegate the signaling arising with entanglement access control to the classical network impose an additional requirement on the strategy, namely, the role assumed by each network node must be locally known at the node, as described in Section~\ref{sec:3}. Furthermore, additional conditions arise with the slotted time organizations on the functional $\chi(\cdot)$. To highlight such a dependency, in the following we denote with $\chi_{T_d}(\cdot)$ the strategy in the downlink, and with $\chi_{T_u}(\cdot)$ the strategy in the uplink.

\section{Entanglement Access Control}
\label{sec:3}

\begin{figure*}[t]
    \centering
    \begin{subfigure}[b]{0.48\textwidth}
        \centering
        \includegraphics[width=\textwidth]{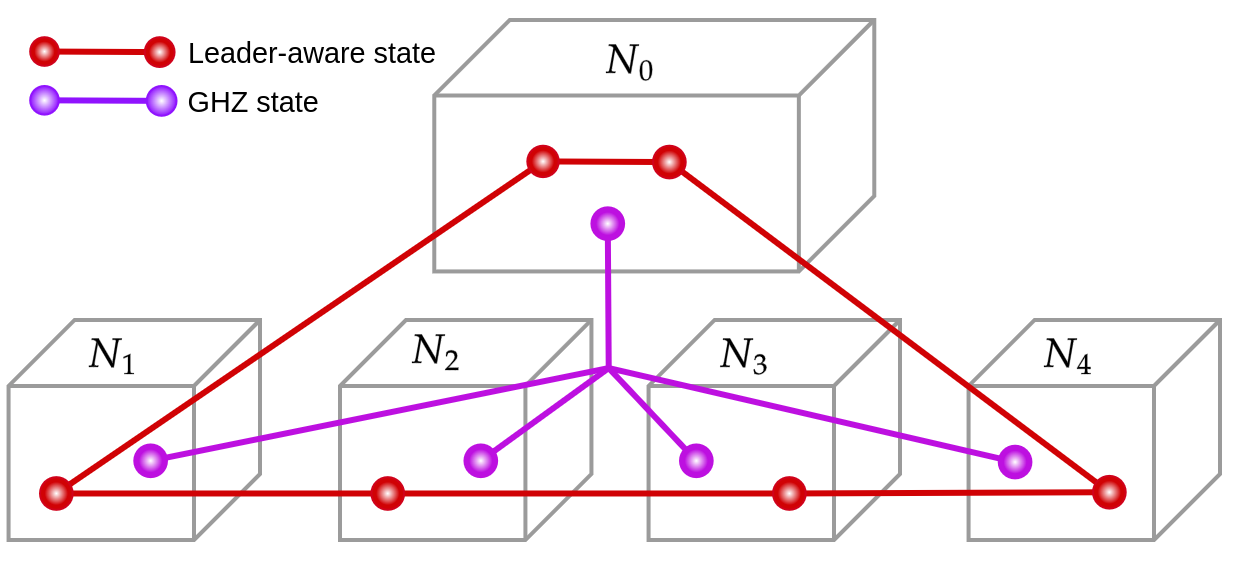}
        \subcaption{Initial state: before contention resolution. The \textit{leader-aware} quantum state enables the contention resolution, whereas the GHZ state supplies the contention winner with the communication resources -- namely, the EPR -- for quantum teleportation.}
        \label{fig:03.a}
    \end{subfigure}
    \hfill
    \begin{subfigure}[b]{0.48\textwidth}
    	\centering
    	\includegraphics[width=\textwidth]{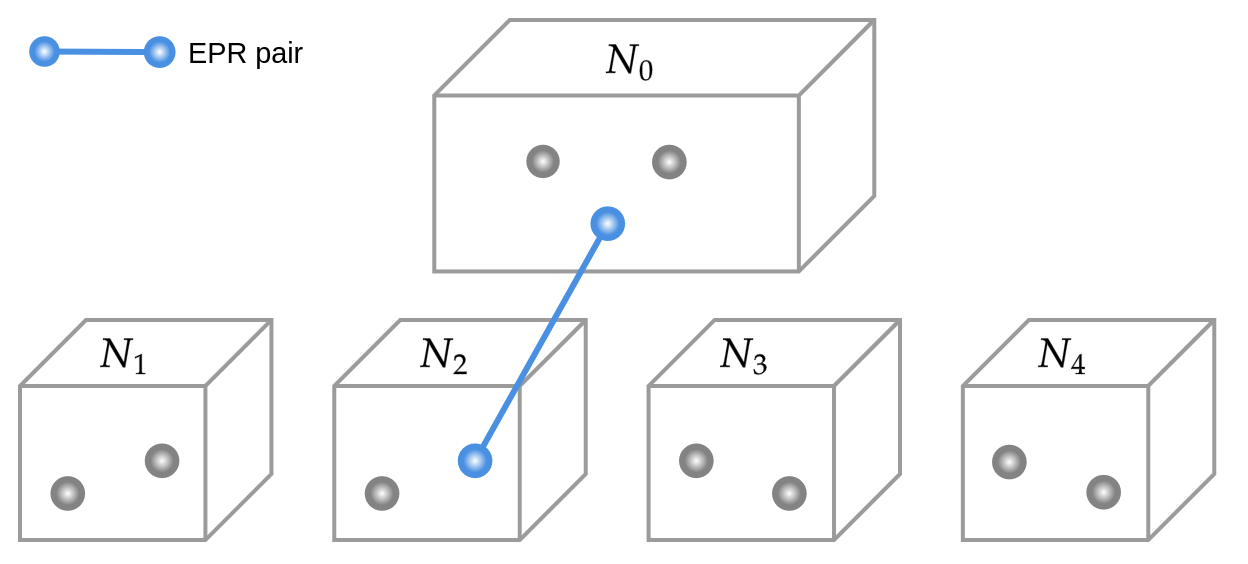}
    	\subcaption{Final state: contention resolution completed. An EPR between the orchestrator $N_0$ and end-node $N_2$ has been generated, with the orchestrator (end-node) acting as transmitter during downlink (uplink) time-slots.}
    	\label{fig:03.b}
    \end{subfigure}
    \caption{Graphical representation of the multipartite quantum states shared among the nodes: (a) once entanglement has been distributed, and (b) once contention has been solved.}
    \label{fig:03}
    \hrulefill
\end{figure*}

Among the different classes of multipartite entangled states \cite{NieChu-11}, two classes exhibit astonishing properties: \textit{GHZ} states \cite{GreHorZei-89,GreHorZei-90} and \textit{W} states \cite{DurVidCir-00}. Specifically, we exploit the maximally connectedness property of the former class for fulfilling the first requirement of the problem statement, i.e., the extraction of an EPR pair, and the asymmetry of the latter state for solving the entanglement contention.

Accordingly, the $n+1$-qubit GHZ state -- represented in red in Figure~\ref{fig:03.a} -- represents the key entangled resource for communication to be shared among the nodes. Furthermore, the $n$-qubit W  state -- which is one of the inputs for the generation of the \textit{leader-aware} entangled state
represented in purple in Figure~\ref{fig:03.a} -- represents the key tool for solving the GHZ state contention. Through a joint processing of both the entangled resources, the designed protocol fulfills all requirements described in Section~\ref{sec:2.1}, providing a collision-free solution to the entanglement contention with probability one.

To better introduce the role of each entangled resource, we represent in Figure~\ref{fig:03} the distribution of the entangled states among the set $\mathcal{N}$ of nodes. Specifically, before the start of the contention resolution the nodes share the two entangled resources as depicted in Figure~\ref{fig:03.a}. Then, once the entanglement access control is performed, an EPR pair -- represented in blue in Figure~\ref{fig:03.b} -- is extracted from the GHZ state and assigned according to the mapping $\chi$ in \eqref{eq:2}. Clearly, once the EPR pair is used for communicating, a new entanglement generation and distribution process occurs.

\subsection{Communication Resource}
\label{sec:3.1}

The maximally connectedness property \cite{NieChu-11} of the GHZ state allows the extraction of an EPR pair that is: i) \textit{deterministic}, i.e., an EPR pair is obtained with probability one\footnote{Clearly neglecting the noise introduced by entanglement distribution and quantum operations for the sake of simplicity.}; ii) \textit{invariant with respect to node identities}, i.e., an EPR pair can be extracted between any pair of qubits (hence, nodes) belonging to the multipartite state; iii) \textit{LOCC}, i.e., the extraction relies exclusively on local operations and classical communications. 

The extraction process works as follows. As said, the input entangled state for EPR generation is a $(n+1)$-GHZ state:
\begin{equation}
    \label{eq:3}
     \ket{GHZ}_{n+1} = \frac{1}{\sqrt{2}} \big( \ket{0}^{\otimes (n+1)}+\ket{1}^{\otimes (n+1)} \big)
 \end{equation}
By recalling the formalism of \cite{EngKraKRa-22},
we denote with $P = \{p_0,..,p_n\}$ the ordered sequence of $n+1$ terms $p_i \in \{0,1\}$, with the subscript $i$ denoting a certain node identity. In order to allow the designed strategy to select a unique transmitter-receiver pair, only two terms in $P$ are equal to one and the remaining are equal to 0. In the next section we clarify that the generation of $P$ with such a property is delegated to the other quantum resource involved in the designed protocol, i.e., the \textit{leader-aware} entangled state, and it is distributed. 

The value of $p_i \in P$ determines the local unitary to be performed on the $i$-th GHZ qubit at the $i$-th node for distributively extracting an EPR pair, as follows:
\begin{align}
    U_{p_i} &= \begin{cases}
        H & \textrm{if } p_i=0\\
        I & \textrm{if } p_i=1\\
    \end{cases}
    \label{eq:4}
\end{align}
with $I$ and $H$ denoting the identity and the Hadamard unitary, respectively. From \eqref{eq:4}, it results that the overall unitary $U_{P}$ acting on the $(n+1)$-GHZ state is given by:
\begin{equation}
    U_{P} = U_{p_0}\otimes U_{p_1} \otimes \ldots \otimes U_{p_n}
    \label{eq:5}
\end{equation}

By applying the operator $U_P$ to state in \eqref{eq:3}, after some algebraic manipulations\footnote{The operator $U_P$ is the result of the tensor product of single-qubit operations. Hence, we can freely rearrange the terms in \eqref{eq:5} so that the first two qubits correspond to the two $p_i$ terms equal to 1.}, the output state is: 
\begin{align}
    \label{eq:6}
    U_P\ket{GHZ}_{n+1} = & \ket{\Phi^+} \otimes \sum_{k=1}^{2^{n-1}}\ket{\psi^{k}_{\text{even}}}_{n-1}\nonumber\\ & + \ket{\Phi^-} \otimes \sum_{k=1}^{2^{n-1}}\ket{\psi^{k}_{\text{odd}}}_{n-1}.
\end{align}
In \eqref{eq:6}, $\ket{\Phi^+}$ and $\ket{\Phi^-}$ denotes the two Bell states resulting from the actions of the two identities. Instead, the $(n-1)$-qubit states $\ket{\psi^{k}_{\text{even}}}_{n-1}$ and $\ket{\psi^{k}_{\text{odd}}}_{n-1}$ denote the states resulting from the processing induced by the $(n-1)$ unitaries $H$. These states, as suggested by the name, are characterized by an \textit{even} and \textit{odd} number of qubits in state $\ket{1}$.

It is worthwhile to note that, by simply measuring the qubit at each \textit{loser} node characterized by $p_i = 0$, namely, a node not selected as transmitter or receiver, an EPR shared between the contention winning nodes is deterministically generated. As a matter of fact, the specific Bell state -- $\ket{\Phi^+}$ or $\ket{\Phi^-}$ --  generated can be determined  by exploiting the properties of states $\ket{\psi^{k}_{\text{even}}}_{n-1}$ and $\ket{\psi^{k}_{\text{odd}}}_{n-1}$. Specifically, a parity check on the measurement outcomes allows to distinguish the two states. 

\subsection{Resource Contention}
\label{sec:3.2}

\textit{W} states allow a fair -- namely, equiprobably -- election of a leader among a set of nodes \cite{Pan-05}. Indeed, this peculiar feature comes straightforward from the state expression:
\begin{equation}
    \label{eq:7}
    \ket{W}_n = \frac{1}{\sqrt{n}} \big( \ket{100\cdots0}+\ket{010\cdots0}+\cdots+\ket{000\cdots1} \big) 
\end{equation}

The election of the leader is fair because of the amplitude of the $\ket{W}_n$ state. Indeed, let us consider a set of $n$ nodes ${N_1, N_2,..,N_n}$, with node $N_i$ detaining the $i$-th qubit -- denoted as $\ket{W_i}$ for the sake of simplicity and with a small notation abuse -- of state $\ket{W}_n$. Each node simply performs a local measurement on its qubit. Accordingly, only one node of the set observes the outcome $1$ -- becoming so the leader -- whereas the remaining nodes observe the outcome $0$. Crucially, each node can observe the outcome $1$ with the same probability equal to $1/n$.  Hence, each node experiences the same probability of being elected leader.

Stemming from a W state, for fulfilling the properties detailed in Section~\ref{sec:2.1}, we enrich the $\ket{W}_n$ state with ancilla qubits. We refer to the global state as the \textit{leader-aware} state $\ket{\Lambda}_{n+m}$, where $m = \lceil\log_2 n\rceil$ is the number of ancilla qubits $\ket{a_1 \ldots a_m}$. Specifically, each end-node stores a qubit of the $\ket{W}_n$ state, whereas the ancilla qubits belongs to the orchestrator\footnote{Although discussed separately, W and ancilla qubits are jointly generated  as a single multi-partite state, and then distributed through the network without the need of CNOTs among remote nodes.}. They are exploited to make the orchestrator aware of the identity of the node elected as receiver (transmitter) during downlink (uplink) time-slots without exchange classical signaling. The circuit for generating the leader-aware state $\ket{\Lambda}_{n+m}$, represented in Figure~\ref{fig:04} for a particular case, can be generalized as a sequence of CNOTS  according to the following rule: for any $\ket{W_i}$, given the binary representation of index $i-1$ equal to $\sum_{j=0}^{m-1} b_j 2^j$, a $\texttt{CX}(\ket{W_i},\ket{a_j})$ with $\ket{W_i}$ acting as control is required for any $b_j \neq 0$.

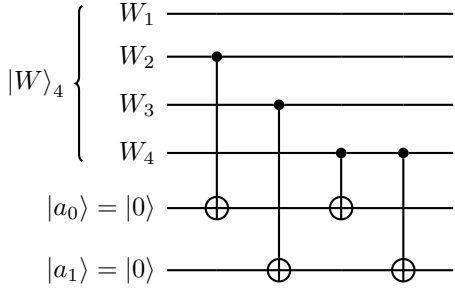
\begin{figure}[t]
    \begin{center}
        \begin{quantikz}
            \lstick[wires=4]{$\ket{W}_4$}& &  \lstick{$W_1$} & \qw & \qw & \qw & \qw & \qw \\
            & & \lstick{$W_2$} & \ctrl{3} & \qw & \qw & \qw & \qw \\
            & & \lstick{$W_3$} & \qw & \ctrl{3} & \qw & \qw & \qw \\
            & & \lstick{$W_4$} & \qw & \qw & \ctrl{1} & \ctrl{2} & \qw \\
            & & \lstick{$\ket{a_0}=\ket{0}$} & \targ{} & \qw & \targ{} & \qw & \qw \\
            & & \lstick{$\ket{a_1}=\ket{0}$} & \qw& \targ{} & \qw & \targ{} & \qw
        \end{quantikz}
    \end{center}
    \caption{Quantum circuit for generating the leader-aware state $\ket{\Lambda}_{6}$ by considering the example in Figure~\ref{fig:03}. The input state of the circuit is $\ket{W}_4 \otimes \ket{00}$, with $\ket{W_i}$ denoting the $i$-th qubit of the $\ket{W}_4$ state. Accordingly $\ket{\Lambda}_6=  \frac{1}{\sqrt{4}}[\ket{100000}+\ket{010010}+\ket{001001}
        +\ket{000111}]_{W_1,W_2,W_3,W_4,a_0,a_1}$.}
    \label{fig:04}
    \hrulefill
\end{figure}


\subsection{Uplink Protocol}
\label{sec:3.3}

We now focus on the entanglement access protocol in the uplink time-slot, during which the orchestrator node $N_0$ acts as receiver. Accordingly, the problem statement in Sec.~\ref{sec:2.2} can be revised as follows:
\begin{align}
    \label{eq:8}
    \mathcal{N_R} = \big\{ \, &  \{(N_i, N_0)\}^n_{i=1} : \nonumber \\
    & \quad N_i \text{ has role } R_t \, \wedge \, N_0 \text{ has role }\overline{R}_t \big\} \\
        \label{eq:9}
    \chi_{T_u} : \, & \mathcal{N_R} \rightarrow \{0,1\} \quad \textrm{ s.t. } \quad \sum_{i} \chi_{T_u} (N_i,N_0) = 1
\end{align}

\begin{table}[b]
    \centering
    \caption{True table of the elected leader discrimination function for the quantum circuit given in Figure~\ref{fig:04}.}
    \begin{tabular}{ | c | c | c | c | | c | c | }
        \hline
        $W_1$ & $W_2$ & $W_3$ & $W_4$ & $a_0$ & $a_1$ \\
        \hline
        1 & 0 & 0 & 0 & 0 & 0 \\
        0 & 1 & 0 & 0 & 1 & 0 \\
        0 & 0 & 1 & 0 & 0 & 1 \\
        0 & 0 & 0 & 1 & 1 & 1 \\
        \hline
    \end{tabular}
    \label{tab:01}
\end{table}

\begin{figure*}[t!]
    \centering
        \includegraphics[width=0.9\textwidth]{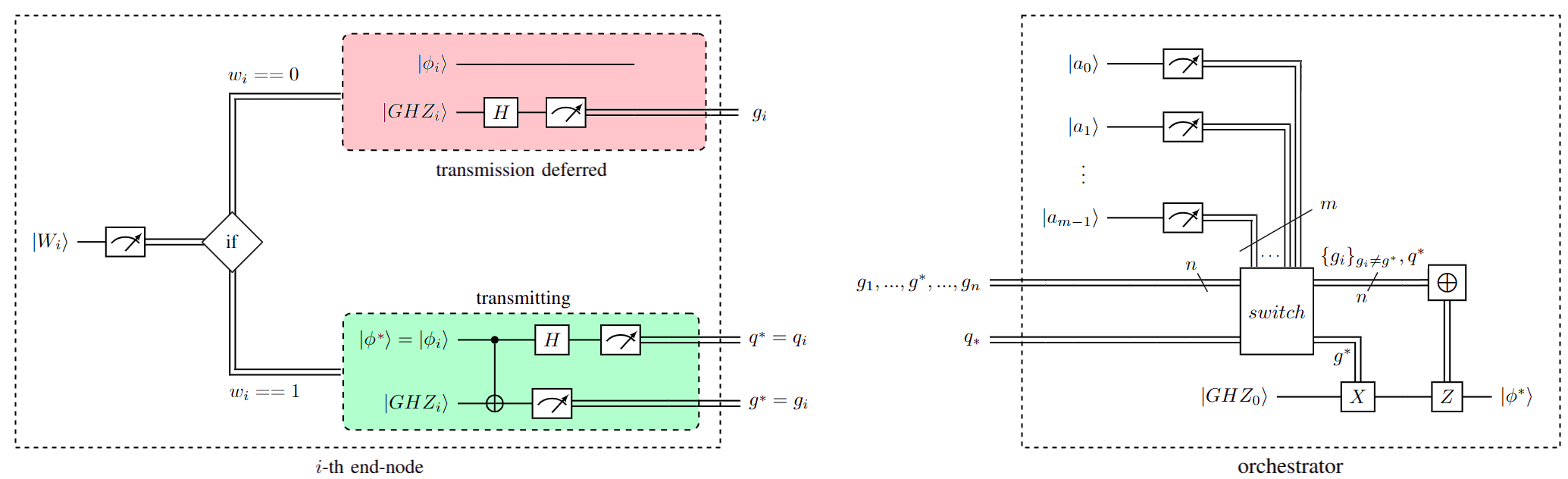}
        \caption{Quantum circuit for uplink entanglement access.}
    \label{fig:05}
    \hrulefill
\end{figure*}

The protocol, represented in Figure~\ref{fig:05}, works as follows. Each end-node $N_i$ performs a local measurement in the computational basis on its qubit $\ket{W_i}$, by obtaining the measurement outcome $w_i \in \{0,1\}$. Such an outcome determines the value of $p_i=w_i$, by univocally determining the unitary in \eqref{eq:4}.

Specifically, whenever $w_i=0$, $N_i$ \textit{loses} the contention -- $\chi_{T_u} (N_i,N_0) = 0$. And consequently $N_i$ performs a local measurement in the Hadamard basis on its qubit $\ket{GHZ_i}$, with measurement outcome denoted as $g_i\in\{0,1\}$. This measurement guarantees $N_i$ to ``leave'' the entangled state $\ket{GHZ}_{n+1}$ by preserving the entanglement shared between the remaining nodes.

Otherwise, whenever $w_i=1$, $N_i$ is granted access to the entanglement resource as transmitter -- $\chi_{T_u} (N_i,N_0) = 1$. Due to the processing performed by the contention losers on their GHZ qubits,  the protocol ends up with the feature of $N_i$ being the only one sharing a ready-to-use EPR pair with the orchestrator as represented in Fig.~\ref{fig:03.b}. Accordingly, it can starts the quantum teleportation as shown in the upper-left part of Figure~\ref{fig:05}. More into detail, $N_i$ performs a CNOT with the qubit to be teleported $\ket{\phi^*}$ acting as controller and the qubit $GHZ_i$ acting as target, followed by an Hadamard gate on $\ket{\phi^*}$ and a measurement on both $\ket{\phi^*}$ and $GHZ_i$ obtaining the outcomes $q*$ and $g^*$, respectively.

From the above, remarkably, the identity of the elected transmitter is not disclosed to the other end-nodes. In fact, the protocol design provides each end-node $N_i$ with only local knowledge about $\chi_{T_u} (\cdot)$ as a result of the measurement of the W qubit locally stored\footnote{The knowledge that may be guessed by observing the transmission of $q_*$ from the winner end-node can be easily obfuscated through dummy random values sent by loser nodes.}. Differently, full knowledge about $\chi_{T_u} (\cdot)$ is available at the orchestrator by simply measuring the ancilla qubits, as shown with the example in Table~\ref{tab:01}.

It is worthwhile to note that the outcomes $\{g_i\}$ from the measurement of the GHZ qubits at the end-nodes must be transmitted to the orchestrator, as shown in Figure~\ref{fig:05}. As a matter of fact, $g^*$ is required for completing the teleportation process \cite{CacCalVan-20} whereas the $\{ g_i \}_{g_i \neq g*}$ are needed for distinguishing between the two Bell states in \eqref{eq:6}. By feeding with all the $\{g_i\}$ a switch controlled by the measurements of the ancilla qubits, the receiver is able to distinguish $g_*$ from $g_i$. Thus the orchestrator is able, by performing a Z gate controlled by the sum modulo 2 of the $\{ g_i \}_{g_i \neq g*}$, to recover $\ket{\phi^+}$ from $\ket{\phi^-}$.

In a nutshell, the designed distributed protocol provides fair, collision-free access to a EPR pair deterministically extracted by a multipartite state. We remark that the collision-free property is independent from the number of connected end-nodes, and it relies on the asymmetry of the entangled state $\ket{W}$.

\subsection{Downlink Protocol}
\label{sec:3.4}

\begin{figure*}
    \centering
    \includegraphics[width=0.9\textwidth]{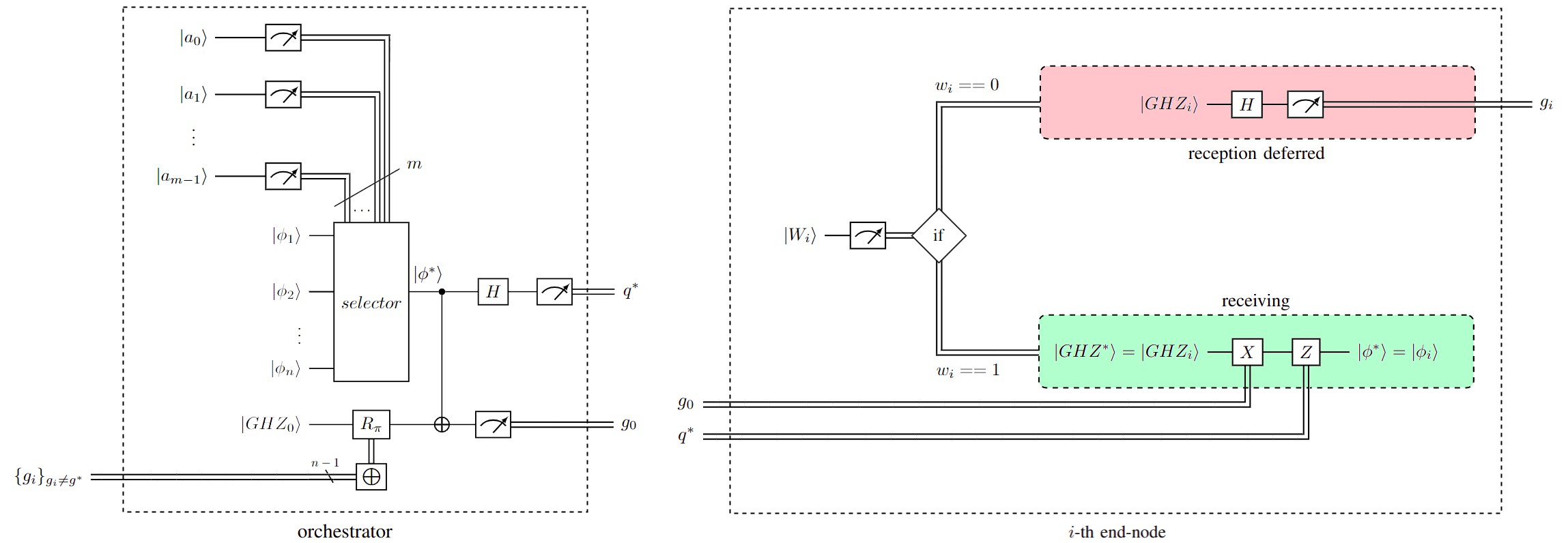}
    \caption{Quantum circuit for downlink entanglement access. }
    \label{fig:06}
    \hrulefill
\end{figure*} 

We now focus on the entanglement access protocol in the downlink time-slot, during which the orchestrator $N_0$ acts
as transmitter. Accordingly, the problem statement becomes:
\begin{align}
    \label{eq:10}
    \mathcal{N_R}= \big\{ \, &  \{(N_0, N_j)\}^n_{j=1} : \nonumber \\
    & \quad N_0 \text{ has role } R_t \, \wedge \, N_j \text{ has role }\overline{R}_t \big\} \\
    \label{eq:11}
    \chi_{T_d} : \, & \mathcal{N_R} \rightarrow \{0,1\} \quad \textrm{ s.t. } \quad \sum_{j} \chi_{T_d} (N_0,N_j) = 1
\end{align}

The circuital representation for the downlink entanglement access is represented in Figure~\ref{fig:06}. Downlink communications represents the dual scenario of uplink communications. More into detail, role $\overline{R_t}$ must be univocally assigned to a single end-node, and the knowledge of such an assignment should be available at the orchestrator for properly scheduling the transmission of the qubit meant to be transmitted to the selected end-node. Due to the problem duality, the same strategy adopted for the assignment and signaling about the transmitter identity in uplink can be used also for the downlink signaling and assignment.

More in details, when $w_i=1$ is measured, node $N_i$ acquires the receiver role, waiting  for the classical bits $g_0$ and $q_*$ needed to complete the teleportation, as shown in lower-left part of Figure~\ref{fig:06}. Otherwise, $N_i$ defers the reception of a downlink communication by leaving the entangled state $\ket{GHZ}_{n+1}$ without altering the entanglement shared between the remaining nodes, as shown in upper-left part of Figure~\ref{fig:06}. As regards to the orchestrator, the ancilla qubits are exploited for selecting the information qubit $\ket{\phi*}$ meant to be transmitted to the end-node that measured $w_i = 1$.

\section{Discussion and Conclusions}
\label{sec:4}
Entanglement access control represents a key functionality of the Quantum Internet. With this work, we proposed a protocol to jointly solve entanglement contention and extraction of an EPR pair from a multipartite entangled state in a distributed way. This result is reached by exploiting the distinctive properties GHZ and W states, obtaining so a collision-free protocol able to incorporate any signaling arising with entanglement access into quantum communication resources.

\bibliographystyle{IEEEtran}
\bibliography{biblio.bib}

\end{document}